\documentclass{emulateapj}
\usepackage{apjfonts}

\shortauthors{R. E. Louis et al.}

\begin{document}

\title{Supersonic Downflows in a Sunspot Light Bridge}

\author{Rohan E. Louis\altaffilmark{1}, Luis R. Bellot Rubio\altaffilmark{2}, 
Shibu K. Mathew\altaffilmark{1} and P. Venkatakrishnan\altaffilmark{1}}

\altaffiltext{1}{Udaipur Solar Observatory, Physical Research Laboratory,
                     Dewali, Badi Road, Udaipur,
	       	     Rajasthan - 313004, India; eugene@prl.res.in}

\altaffiltext{2}{Instituto de Astrof\'{\i}sica de Andaluc\'{\i}a (CSIC),
                     Apartado de Correos 3004,
                     18080 Granada, Spain}

\begin{abstract}
We report the discovery of supersonic downflows in a sunspot light
bridge using measurements taken with the spectropolarimeter on board
the {\em Hinode} satellite. The downflows occur in small patches close
to regions where the vector magnetic field changes orientation
rapidly, and are associated with anomalous circular polarization
profiles. An inversion of the observed Stokes spectra reveals
velocities of up to 10~km~s$^{-1}$, making them the strongest
photospheric flows ever measured in light bridges. Some (but not all)
of the downflowing patches are cospatial and cotemporal with
brightness enhancements in chromospheric \ion{Ca}{2} H filtergrams. We
suggest that these flows are due to magnetic reconnection in the upper
photosphere/lower chromosphere, although other mechanisms cannot be
ruled out.

\end{abstract}


\section{Introduction}
\label{intro}

Light bridges (LBs) are bright structures in the otherwise dark 
umbra that often exhibit a granular-like morphology
\citep{Muller79,Sobt94,Hirzb02,Rim08}. They represent a discontinuity or
interruption in the regular umbral field \citep{Gar87}. LBs are known
to harbor weak and inclined fields \citep{Rud95,Leka97,Jur06,Kat07a},
but their origin and magnetic nature is still a matter of debate
\citep{Park79,Choud86,Spruit06,Rim97,Rim04}.

One of the reasons that make LBs interesting is their chromospheric
activity, seen as surges in H$\alpha$ \citep{Asai01}, localized
brightenings in \ion{Ca}{2} H filtergrams \citep{Rohan08, Shimizu09},
and brightness enhancements in the upper chromosphere and transition
region \citep{Berger03}. The stressed magnetic configuration
associated with LBs is perceived to be responsible for this 
activity, but the precise relationship, if any, is yet to be
established. Unfortunately, there is a lack of simultaneous vector
magnetic field measurements in the photosphere to diagnose these
events.

Here we analyze {\em Hinode} spectropolarimetric observations and
\ion{Ca}{2} H filtergrams of a sunspot LB in an attempt to relate 
its chromospheric activity to the photospheric magnetic field. We
discover patches of supersonic downflows in the photospheric layers 
of the LB and show that some of them are associated with strong
\ion{Ca}{2} H brightness enhancements. Interestingly, the
supersonic flows produce spectral signatures never seen before
in LBs.

\section{Observations}
\label{data}
On 2007 May 1, the leading spot of NOAA Active Region 10953 was
observed with {\em Hinode} \citep{Kosugi07} at a heliocentric angle of
8$^\circ$ ($\mu=0.99$). Between 10:46 and 12:25 UT, the {\em Hinode}
spectropolarimeter \citep{Lites01,Ichimoto08,Tsun07} recorded the four
Stokes profiles of the iron lines at 630 nm with a spectral sampling
of 21.55 m\AA\/, a pixel size of 0\farcs16, and an exposure time of
4.8~s per slit position (normal map mode). The observations were
corrected for dark current, flat field, thermal flexures, and
instrumental polarization using routines included in the SolarSoft
package. Simultaneously, the Broadband Filter Imager of {\em Hinode}
took \ion{Ca}{2} H filtergrams with a cadence of 1 minute to monitor
the chromosphere of the LB. The effective pixel size of the Ca images
is 0\farcs11.

\begin{figure*}[t]
\centerline{
\includegraphics[angle=90,scale=0.348,bb= 25 275 588 857]{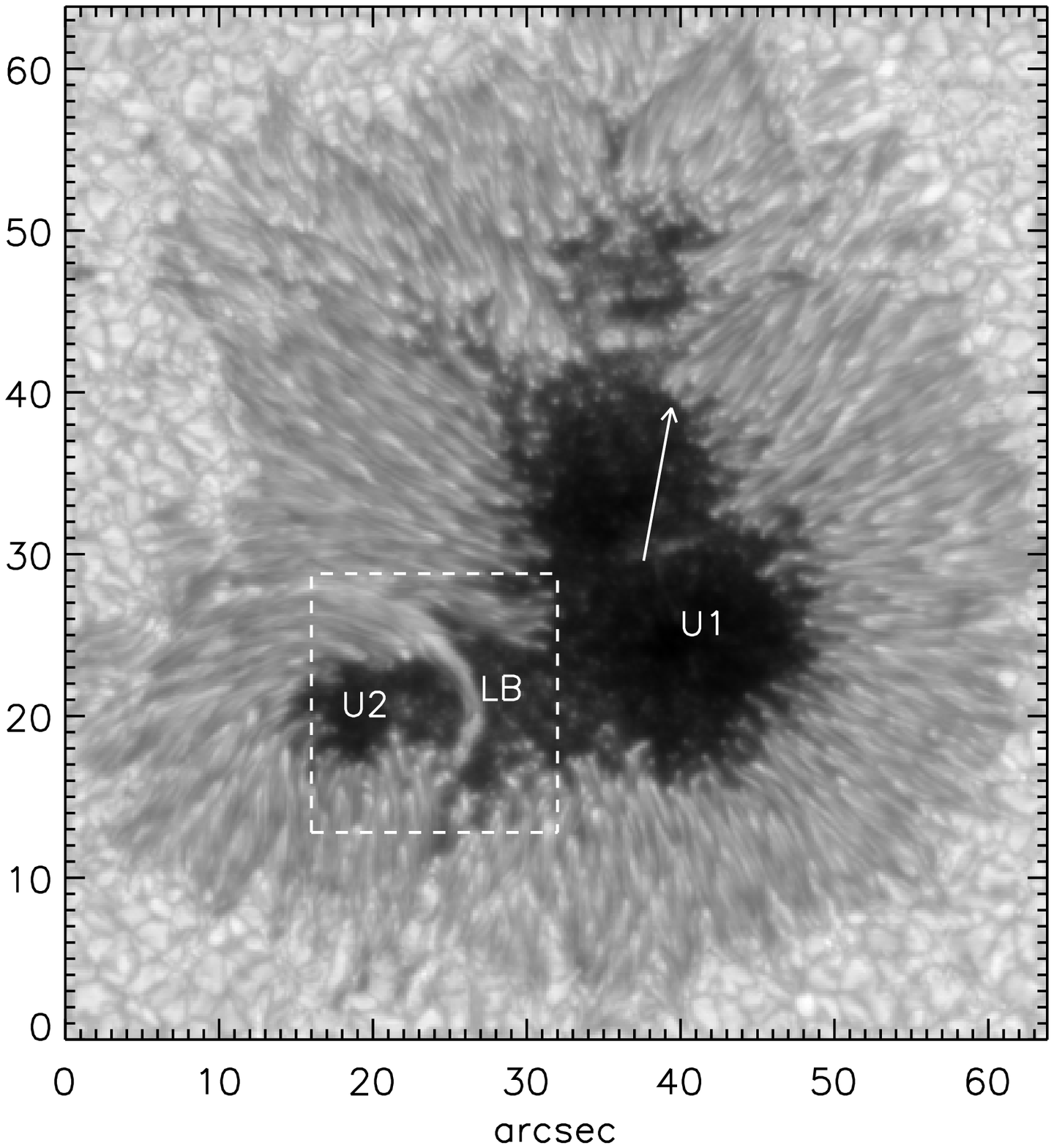}
\includegraphics[angle=90,scale=0.348,bb= 25 170 588 492]{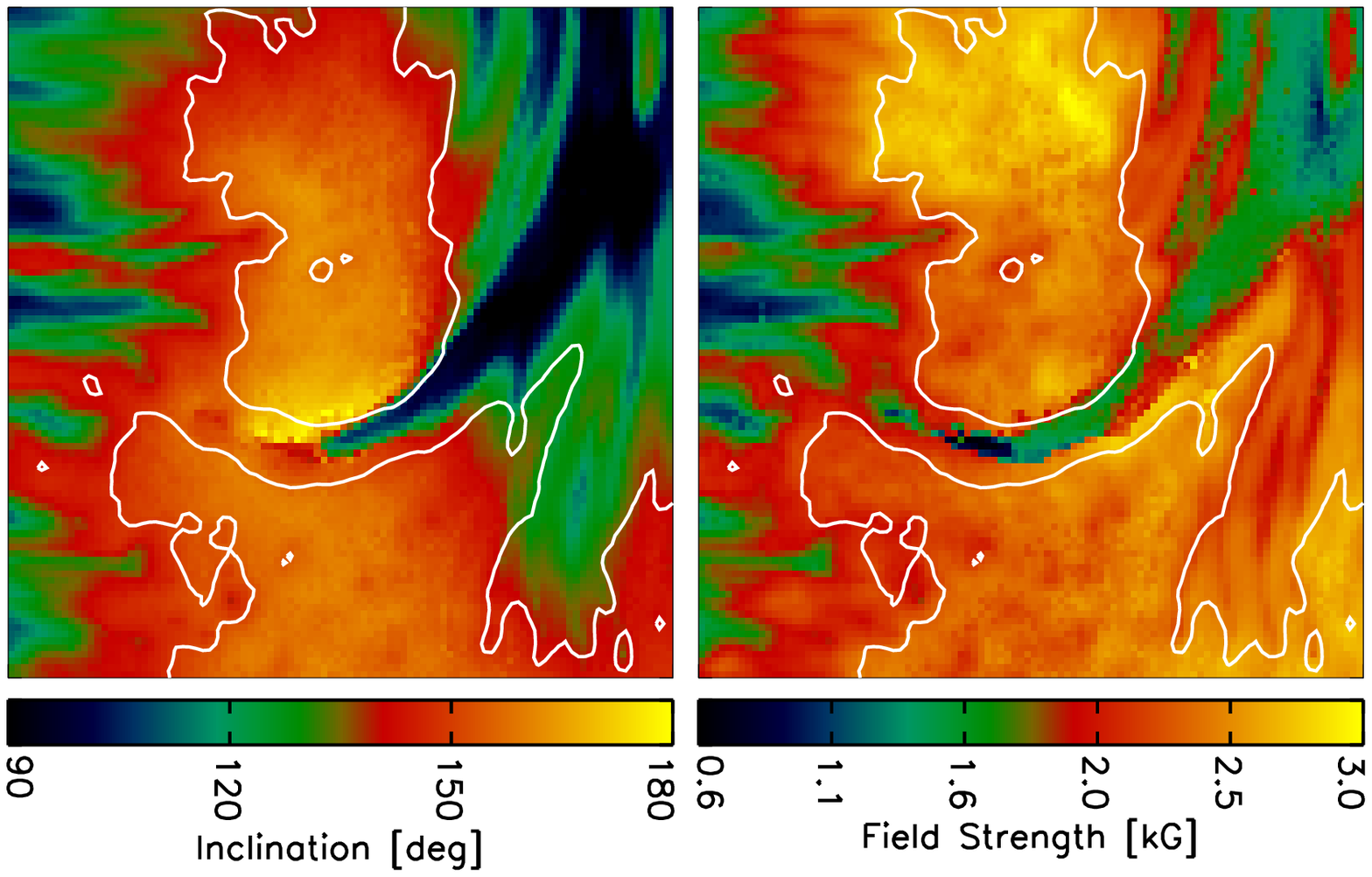}
\includegraphics[angle=90,scale=0.348,bb= 25 170 588 492]{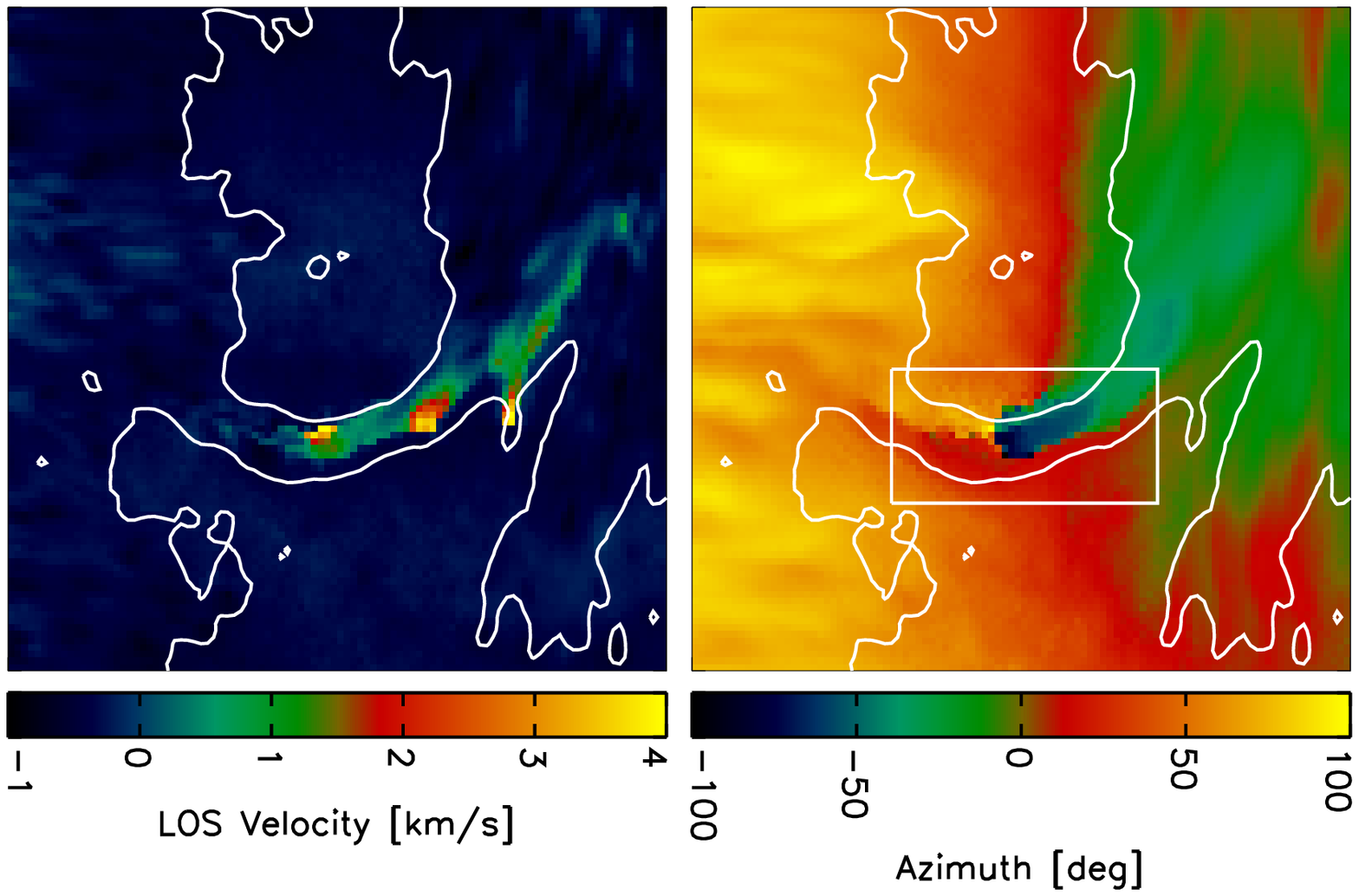}
\includegraphics[angle=90,scale=0.348,bb= 25 180 588 502]{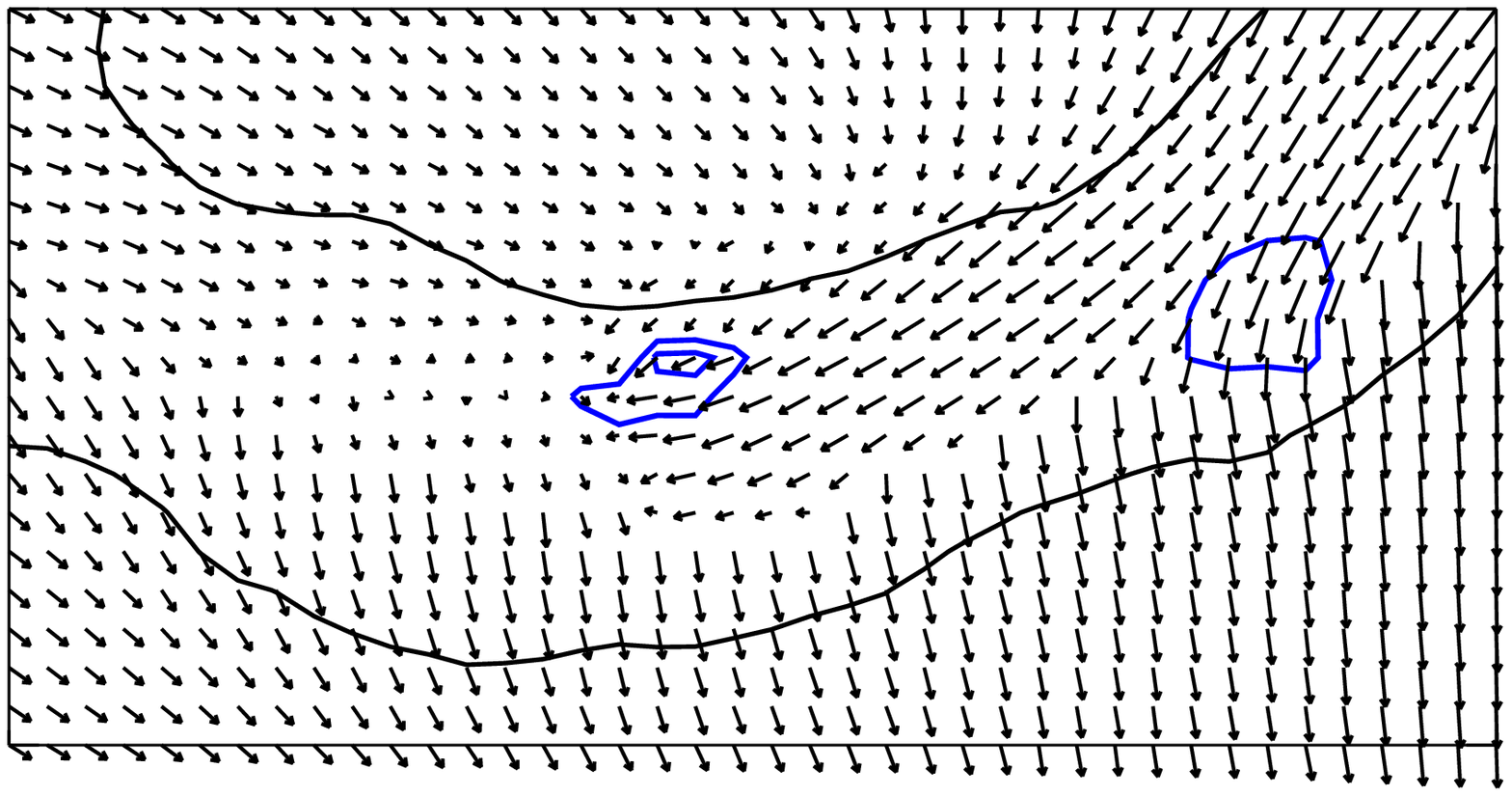}
}
\caption{{\em First column}: Continuum image of NOAA AR 10953 at 630~nm. The 
white box represents the region containing the light bridge (LB), while U1 and
U2 are the sunspot umbrae adjacent to the LB. North is up and West to
the right. The arrow indicates the direction to disk center.  {\em
Second and third columns:} Magnetic field strength, field inclination,
field azimuth, and LOS velocity in the 16$''\times$16$''$ sub-region
containing the LB, as deduced from the inversion. The angles are
expressed in the local reference frame. Azimuths increase
counterclockwise, with zero representing fields pointing to solar
West. Positive velocities indicate redshifts. The mean umbral velocity
is zero. {\em Fourth column}: Transverse component of the vector
magnetic field in the LB. The blue contour lines mark LOS velocities
of 2.5 and 4 km~s$^{-1}$.}
\vspace*{1em}
\label{sunspot_image}
\end{figure*}

\section{Results}
\label{results}

\subsection{Downflowing Patches and Chromospheric Activity}
\label{SIR}

The left panel of Figure~\ref{sunspot_image} shows a continuum map of
the spot and the LB. We have inverted the observed Stokes profiles
using the SIR code \citep[Stokes Inversion based on Response
Functions;][]{Ruiz92}. SIR computes perturbations in the physical
quantities at specific locations across the optical depth grid called
{\em nodes}, and then carries out an interpolation to yield values at
all grid points. To determine the global structure of the LB and the
surroundings, we performed a one-component inversion setting the
magnetic and dynamic parameters to be constant with depth. The
temperature stratification was perturbed with two nodes. A total 
of 9 parameters were retrieved from the observed profiles, including
height-independent micro- and macro-turbulent velocities and a
stray-light factor.

The three components of the vector magnetic field (strength,
inclination, and azimuth) deduced from the inversion are shown in the
second and third columns of Figure~\ref{sunspot_image}. All the angles
are expressed in the local reference frame after a manual 
disambiguation of the line-of-sight (LOS) azimuths. As can be seen,
the LB is characterized by weaker and more inclined fields than the
umbra. This confirms earlier results by, e.g., \cite{Leka97} and
\citet{Jur06}. In the upper half of the LB, the magnetic field is
parallel to the axis of the bridge. Both photometrically and
magnetically, the LB looks like an extension of the penumbra
protruding into the umbra. \cite{Rohan08} detected a horizontal flow
along the LB that starts in the adjacent penumbra, demonstrating that
the two structures are also connected dynamically. At the lower end of
the LB, where the LB fields pointing south encounter sunspot fields
oriented toward the north, one observes an isolated region with
relatively weak magnetic fields. In addition, there is a discontinuity
in the field azimuth running parallel to the west edge of the LB.

\begin{figure}
\begin{center}
\includegraphics[width=0.44\textwidth,angle=90,bb=64 244 590 780]{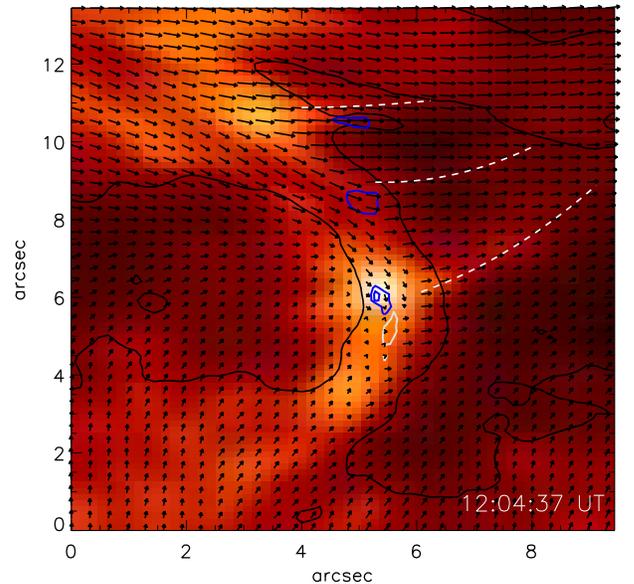}
\end{center}
\caption{\ion{Ca}{2} H filtergram of the LB taken at 12:04:37 UT when 
the spectrograph slit was above the LB. Contours corresponding to LOS 
velocities of 2.5 and 4 km s$^{-1}$ are shown in blue. The 
arrows indicate the transverse component of the field in the local 
reference frame for every alternate pixel, after solving the 
180$^\circ$ azimuth ambiguity. White and black contours represent 
fields weaker than 750 G and bright continuum structures, respectively.
The white dashed lines mark thin chromospheric threads whose ends 
are located in or near the downflowing patches. }
\label{calcium}
\end{figure}

The LOS velocity map displayed in the third column of
Figure~\ref{sunspot_image} reveals the existence of strong, localized
downflows in the LB with velocities of up to 4
km~s$^{-1}$. Interestingly, the downflows occur close to the
weak-field region and the azimuth discontinuity described above, i.e.,
at positions where the magnetic field changes orientation very rapidly
(fourth column of Figure~\ref{sunspot_image}). Some of the downflowing
patches coincide with chromospheric \ion{Ca}{2} H brightness
enhancements, as can be seen in Figure~\ref{calcium}. The filtergram
displayed there was taken during the polarimetric scan of the LB and
shows a strong \ion{Ca}{2} H line-core brightening at the position and
time of the largest photospheric velocities\footnote{The other two
patches do not exhibit significant brightness enhancements. However,
they are located near the end of thread-like chromospheric structures
that reach into the umbra U1 (see the white lines in Figure
\ref{calcium}). These structures show brightenings, but not as intense
as those associated with the strongest downflows.}. NOAA AR 10953
produced many other long-lasting chromospheric plasma ejections on
April 29 and 30 \citep{Shimizu09}.

\begin{figure*}
\begin{center}
\includegraphics[height=10cm,angle=90,bb=20 50 564 852]{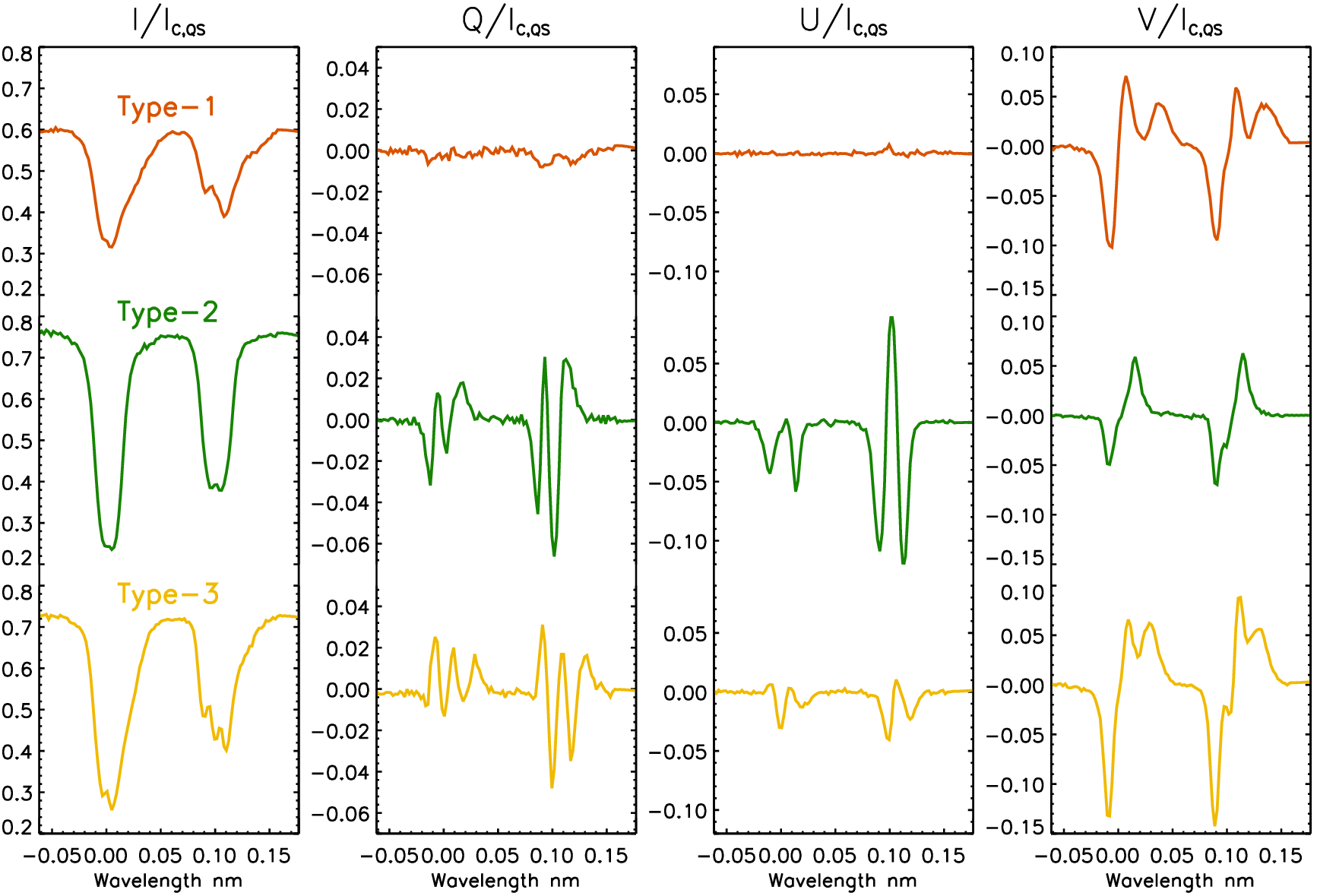} 
\includegraphics[height=6.5cm,angle=90,bb=20 370 650 948]{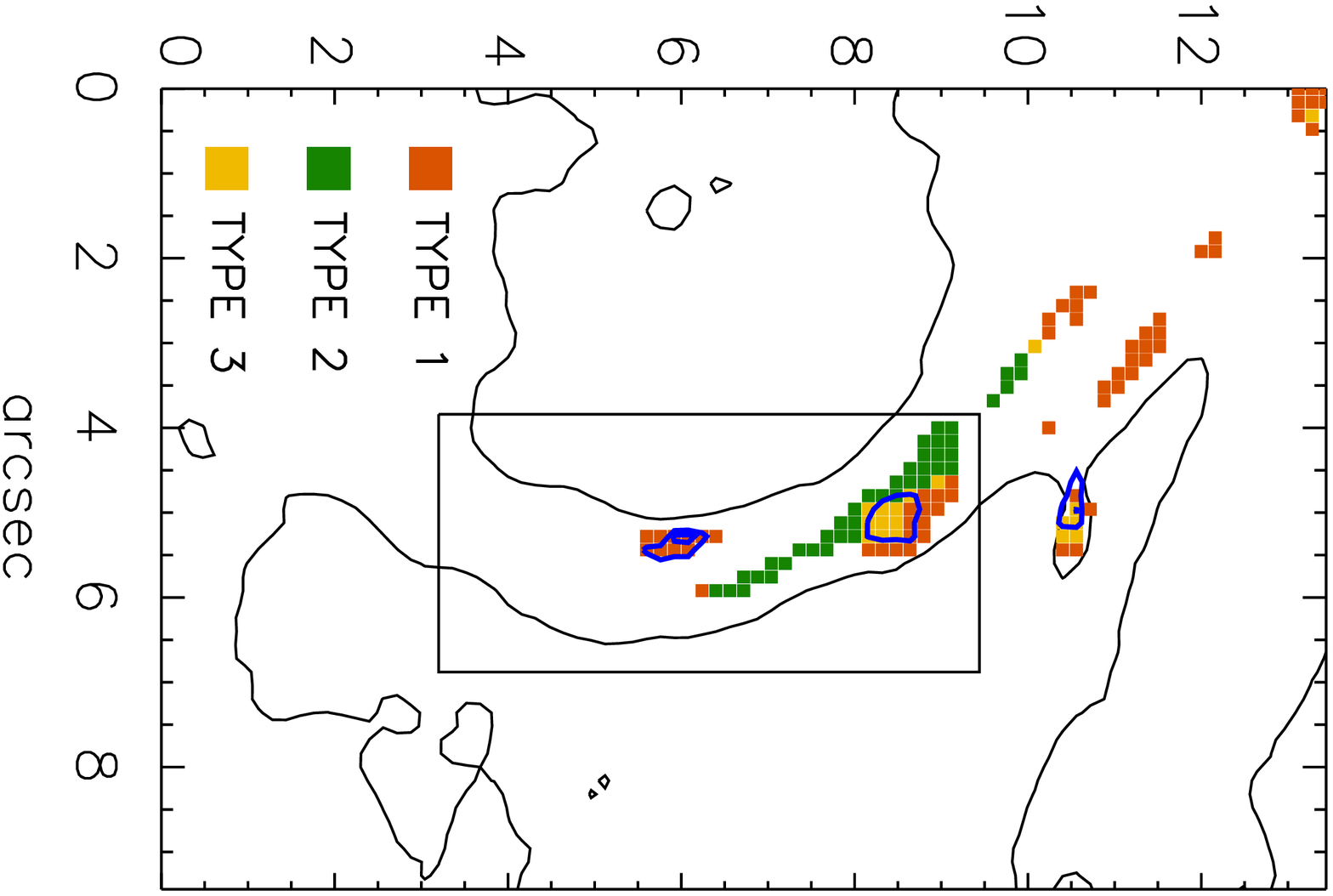}
\end{center}
\caption{{\em Left:} Anomalous Stokes spectra observed in the LB. Type I, 
II, and III profiles are shown in orange, green and yellow,
respectively. {\em Right:} Spatial distribution of the anomalous
profiles. Blue contours mark velocities larger than 2.5 and 4~km~s$^{-1}$.  }
\label{combo_plots}
\end{figure*}

The Stokes $V$ profiles associated with the downflows have two peaks
in the red lobe, i.e., they exhibit a total of three peaks. Hereafter
they will be labelled as Type 1. In the LB one also finds anomalous
linear polarization profiles with normal Stokes $V$ signals which are
designated as Type 2. Type 3 profiles are essentially a combination 
of the other two classes. Examples of these profiles are given in
Figure~\ref{combo_plots}, together with their spatial distribution.

\subsection{Anomalous Stokes $V$ profiles}
\label{sir-inv}
Milne-Eddington-like atmospheres such as the ones used to determine the 
global structure of the LB cannot reproduce the complex shapes of Type 
1 profiles. For this reason, the velocities given in Section~\ref{SIR} 
are only approximate. Here we obtain more reliable values with the help 
of two-component inversions. 

\begin{figure*}
\includegraphics[width=0.37\textwidth,angle=90]{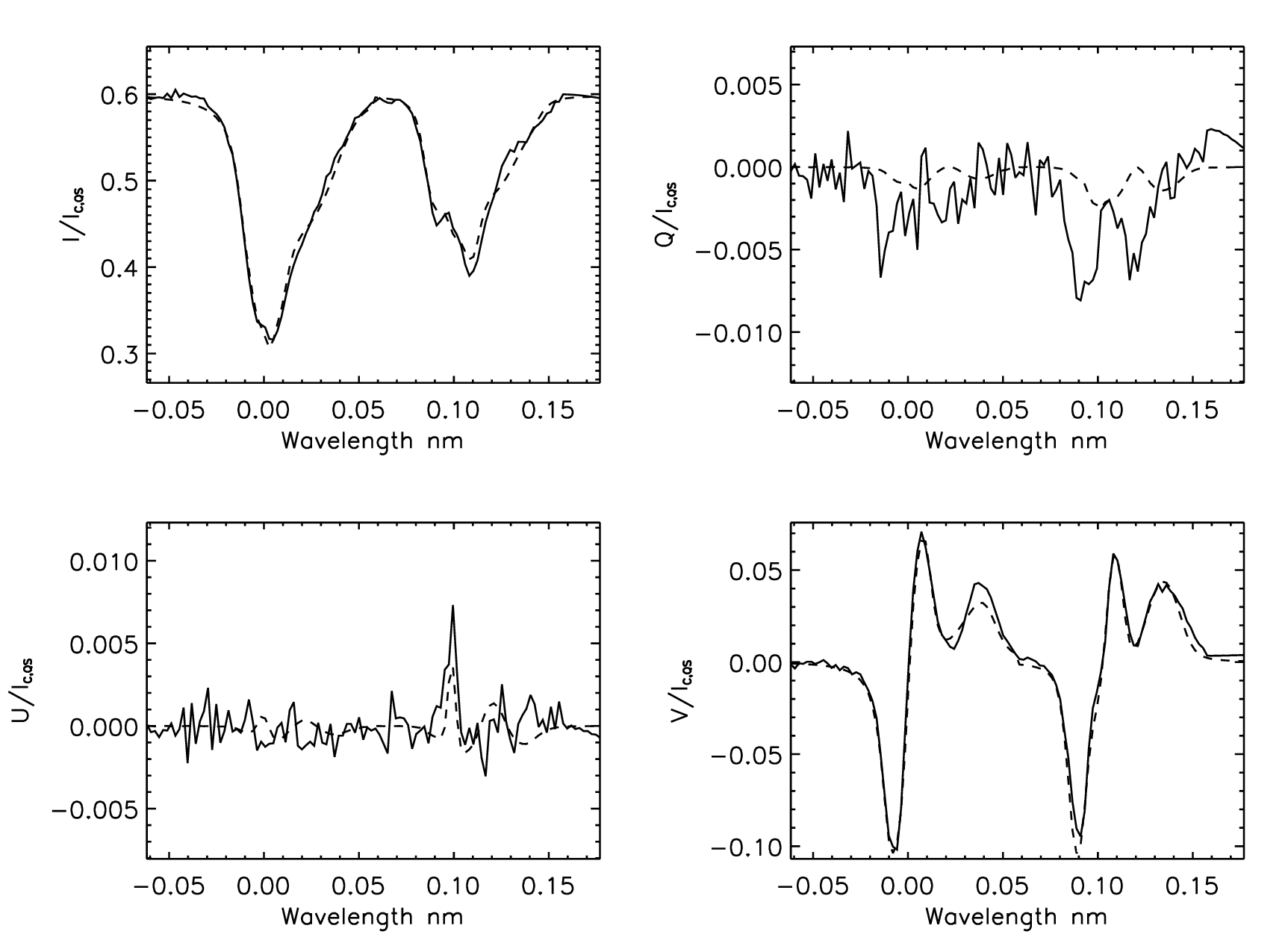}
\includegraphics[width=0.37\textwidth,angle=90]{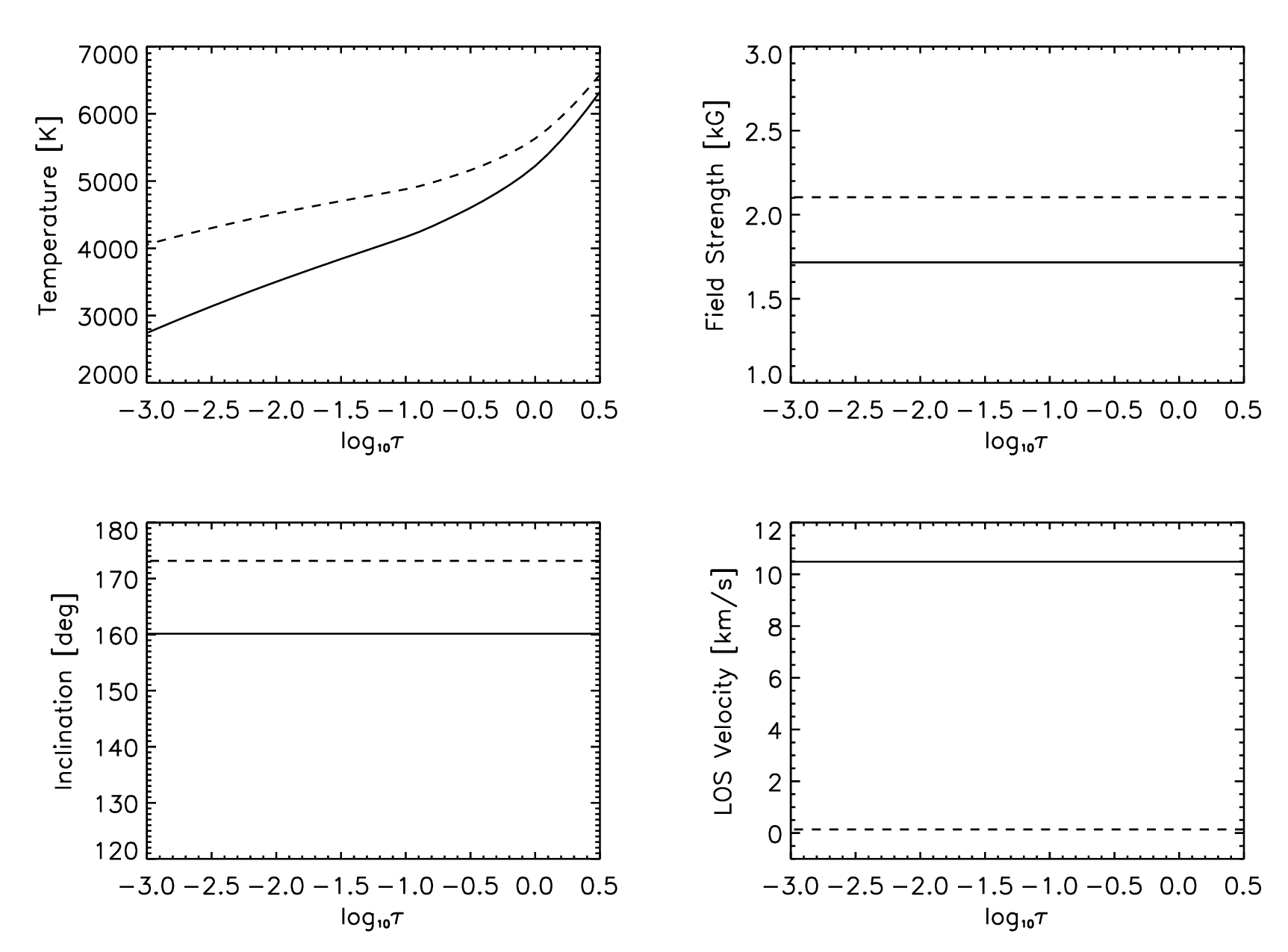}

\caption{{\em First and second columns:} Observed ({\em{solid}}) and best-fit 
({\em{dashed}}) Stokes profiles using a simple two-component atmosphere. 
{\em Third and fourth columns:} Atmospheric stratifications for the two 
components (solid and dashed lines, respectively). The filling factor of 
the fast component is 43\%.}
\label{inv_results1}
\end{figure*}

\begin{figure*}
\includegraphics[width=0.37\textwidth,angle=90]{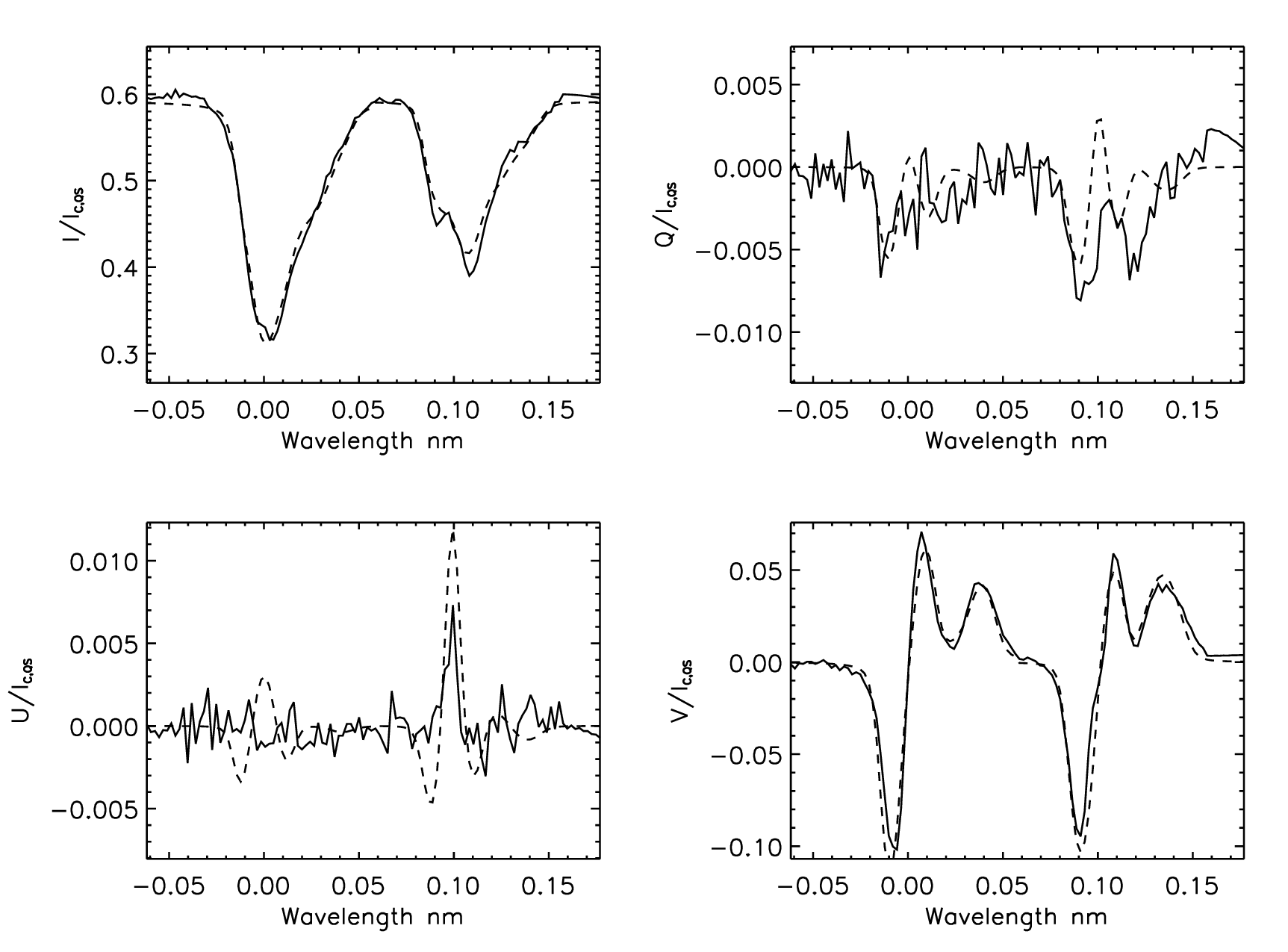}
\includegraphics[width=0.37\textwidth,angle=90]{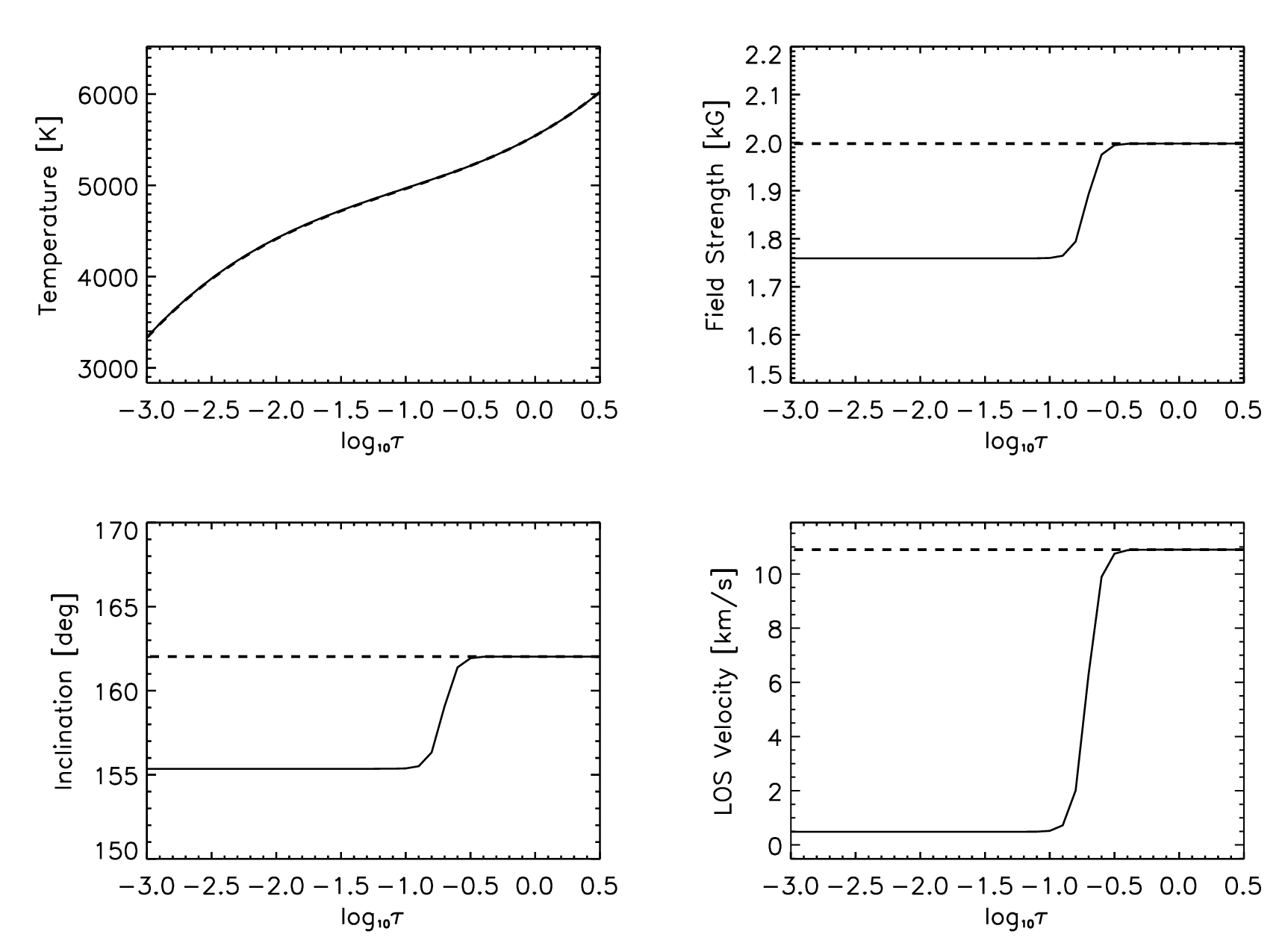}
\caption{Same as Figure~\ref{inv_results1}, for a two-component model in which
one of the component has a discontinuous stratification with
supersonic speeds in the deeper layers. In the third and fourth columns, 
the constant background atmosphere is represented by the dashed lines, 
whereas the solid lines correspond to the discontinuous atmosphere. The latter 
fills 93\% of the resolution element.}
\label{inv_results2}
\end{figure*}

We start with a simple two-component model in which the magnetic field
and the LOS velocity do not vary with height. The results of this 
inversion are presented in Figure 4. As can be seen, one of the 
components has supersonic velocities of $\sim 10$ km~s$^{-1}$ and 
the other is essentially at rest. The model explains the observed 
profiles in a very satisfactory way, but the assumption of two 
magnetic components separated horizontally in the resolution 
element may not be realistic because of the relatively large 
sizes of the downwflowing patches (0.5-0.7 arcsec).
An alternative scenario is that they are stacked in the 
vertical direction, one on top of the other.

To investigate this possibility we consider a different model
atmosphere with two magnetic components. One of them has
height-independent parameters and the other features a discontinuity
in the stratifications at a certain optical depth. The amplitude and
the location of the discontinuity are free parameters.  These
inversions have been carried out using SIRJUMP, an extension of the
SIR code (Bellot Rubio, in preparation). The results show that the
fast component always has velocities exceeding 10 km~s$^{-1}$. As an
example, Figure~\ref{inv_results2} displays the stratifications of 
one such model with the downflows occurring in the lower half of 
the photosphere.

While the two-component models used in this Section differ in complexity,
all of them indicate supersonic velocities and magnetic fields of
the same polarity. In fact, the existence of supersonic flows is
obvious from the shapes of the observed Stokes $V$ profiles, which are
similar to those emerging from the outer part of the penumbra where
the Evershed flow exceeds the sound speed. The inclined red wings of
Stokes $I$ also demonstrate the occurrence of strong velocities. All
these spectral signatures are well reproduced by the two-component
inversions. To the best of our knowledge, this is the first time that
supersonic downflows are found in LBs. Subsonic speeds of $\sim$1
km~s$^{-1}$ have been reported earlier by \citet{Rud95},
\citet{Schl03}, and \cite{Bhartietal2007}.

\subsection{Anomalous Linear Polarization Profiles}
\label{linear-stokes}
In regions where the magnetic field is relatively horizontal, the 
supersonic flows are associated with Type 3 profiles. They have similar
Stokes $I$ and $V$ spectra as Type 1 profiles, but also complex linear
polarization signals which cannot be seen when the field is more
vertical. We observe anomalous Stokes $Q$ profiles with up to five
peaks and signatures of strong redshifts (Figure~\ref{combo_plots},
left). The multiple peaks indicate that different magnetic components
coexist in the resolution element at the position of the downflows.

Type 3 profiles are common near the azimuth discontinuity on the west
side of the LB, where the LB fields meet the umbral fields
(Figure~\ref{combo_plots}, right). Thus, it is likely that these are
the magnetic components giving rise to the complex Stokes $Q$
signals. They may lie horizontally next to each other in the
resolution element, or could be stacked one on top of the other. In
both cases the large variations of field azimuth that occur on very
small spatial scales should facilitate the reconnection of magnetic
field lines from the LB and the adjacent umbra, even if they have the
same polarity \citep{ryutova03}. The interaction is expected to
produce bidirectional jets, and we speculate that the supersonic
downflows observed near the azimuth discontinuity and the weak-field
region are the signatures of reconnection jets directed toward the
photosphere. This would agree with \cite{Shimizu09}, who believe that
the LB is essentially a horizontal, twisted flux tube that reconnects
with the umbral field at a low altitude.
However, our measurements cannot confirm the reconnection scenario 
unambiguously because the upward jet components are not always
visible in the \ion{Ca}{2} H images. Indeed, another
possibility is that the supersonic downflows are produced by a
mechanism similar to that driving the Evershed flow at the outer end
of penumbral filaments.

\section{Summary and Conclusions}
\label{summary}

We have determined the global magnetic and dynamic properties of a
sunspot light bridge from the inversion of Stokes profiles observed by
{\em Hinode}. A one-component model atmosphere with height-independent
parameters (except for the temperature) has been used to that end. Our
analysis reveals patches of strong downflows exceeding 4~km~s$^{-1}$
in the LB, associated with complex Stokes $V$ spectra having
double red lobes. Two-component inversions of the profiles indicate
supersonic velocities of 10 km~s$^{-1}$ or more. A Milne-Eddington 
inversion by \citet{Shimizu09} of the same LB on the previous day 
shows only a small downflow patch with velocities of 0.7 km~s$^{-1}$.

The supersonic downflows occur in regions where the magnetic field of
the LB meets sunspot fields with rather different orientations. In
those locations one observes anomalous linear polarization signals,
but only when the field is relatively horizontal. Moreover, the
supersonic downflows are sometimes associated with transient
chromospheric brightenings. The interaction of the LB field and the
sunspot fields may create current sheets, leading to magnetic
reconnection in the upper photosphere/lower chromosphere. The strong
photospheric flows could represent the downward directed jets
generated in this process. However, such an explanation remains
speculative because the associated upward jets are not always seen 
as chromospheric brightenings.

The discovery of supersonic downflows in LBs has been possible due to
the high spatial resolution and stability of {\em Hinode}, but their
interpretation is still unclear. Additional observations should be
performed to investigate the chromosphere and photosphere of LBs 
simultaneously, in an attempt to detect small brightness enhancements 
that would confirm the reconnection scenario or favor alternative 
explanations.

\vspace{-5pt}

\acknowledgments
Our sincere thanks to the {\em Hinode} team for providing the data
used in this Letter, and to the referee for valuable suggestions. Hinode
is a Japanese mission developed and launched by ISAS/JAXA, with NAOJ
as domestic partner and NASA and STFC (UK) as international
partners. It is operated by these agencies in co-operation with ESA
and NSC (Norway). We acknwoledge support by the Spanish
MICINN through projects ESP2006-13030-C06-02 and PCI2006-A7-0624, and
by Junta de Andaluc\'{\i}a through project P07-TEP-2687.

\end{document}